# COVIDX-Net: A Framework of Deep Learning Classifiers to Diagnose COVID-19 in X-Ray Images

Ezz El-Din Hemdan[1]  , Marwa A. Shouman[1]  , and Mohamed Esmail Karar [2,3]  , *IEEE, Member*

[1]*Department of Computer Science and Engineering, Faculty of Electronic Engineering, Menoufia University, Egypt.*
[2]*Department of Industrial Electronics and Control Engineering, Faculty of Electronic Engineering, Menoufia University, Egypt.*
[3]*Department of Computer Engineering, College of Computing and Information Technology, Shaqra University, Saudi Arabia.*

*EzzElDinHemdan@el-eng.menofia.edu.eg*; *marwa.shouman@el-eng.menofia.edu.eg*; *mekarar@ieee.org*

**Abstract**

**Background and Purpose:** Coronaviruses (CoV) are perilous viruses that may cause Severe Acute Respiratory Syndrome (SARS-CoV), Middle East Respiratory Syndrome (MERS-CoV). The novel 2019 Coronavirus disease (COVID-19) was discovered as a novel disease pneumonia in the city of Wuhan, China at the end of 2019. Now, it becomes a Coronavirus outbreak around the world, the number of infected people and deaths are increasing rapidly every day according to the updated reports of the World Health Organization (WHO). Therefore, the aim of this article is to introduce a new deep learning framework; namely COVIDX-Net to assist radiologists to automatically diagnose COVID-19 in X-ray images.

**Materials and Methods:** Due to the lack of public COVID-19 datasets, the study is validated on 50 Chest X-ray images with 25 confirmed positive COVID-19 cases. The COVIDX-Net includes seven different architectures of deep convolutional neural network models, such as modified Visual Geometry Group Network (VGG19) and the second version of Google MobileNet. Each deep neural network model is able to analyze the normalized intensities of the X-ray image to classify the patient status either negative or positive COVID-19 case.

**Results:** Experiments and evaluation of the COVIDX-Net have been successfully done based on 80-20% of X-ray images for the model training and testing phases, respectively. The VGG19 and Dense Convolutional Network (DenseNet) models showed a good and similar performance of automated COVID-19 classification with f1-scores of 0.89 and 0.91 for normal and COVID-19, respectively. The worst classification performance is obtained for the InceptionV3 model with f1-scores of 0.67 for normal cases and 0.00 for COVID-19 cases.

**Conclusions:** This study demonstrated the useful application of deep learning models to classify COVID-19 in X-ray images based on the proposed COVIDX-Net framework. Clinical studies are the next milestone of this research work.

## 1. Introduction

Coronaviruses (CoV) are a large family of perilous viruses [1]. The CoV is so named because of their characteristic solar corona (crown-like) appearance when observed under an electron microscope [2]. They may cause severe and infectious diseases such as Severe Acute Respiratory Syndrome (SARS-CoV) and Middle East Respiratory Syndrome (MERS-CoV). The outbreak of the 2019 novel coronavirus in Wuhan, China has been rapidly spread to other countries since December 2019 [3-6]. The World Health Organization (WHO) named the infectious disease caused by this kind of viruses as COVID-19 on Feb 11, 2020 [7]. There have been 80,894 confirmed cases to China to date (Mar 18, 2020), and 204,037 confirmed cases worldwide [8].

Although the real-time polymerase chain reaction (RT-PCR) assay of the sputum is the gold standard for Coronaviruses diagnosis, it is time-consuming to confirm COVID-19 patients because of resulted in high false-negative levels [9]. Therefore, medical imaging modalities such as Chest X-ray (CXR) and Computed Tomography (CT) can play a major role in confirming positive COVID-19 patients, especially in cases of infected pregnant women and children [10, 11]. Volumetric CT thorax images for lung and soft tissue have been investigated in the recent studies for identifying COVID-19 [10, 12]. However, the main disadvantage of using CT imaging is the high patient dose and costs scan [13]. In contrast, conventional radiograph or CXR machines are available in all hospitals and clinics to produce 2-dimensional (2D) projection images of the patient's thorax. Generally, the CXR modality is the first choice for radiologists to detect the chest pathology and has been applied to identify or confirm COVID-19 in a small number of patients [10, 14]. Therefore, the focus of this study is only on the use of X-ray imaging modality for potential COVID-19 patients.

Nevertheless, X-ray images cannot easily distinguish soft tissue with a poor contrast to limit the exposure dose to the patients [13, 15]. To overcome these limitations, Computer-Aided Diagnosis (CAD) systems have been developed to assist physicians to automatically detect and quantify suspected diseases of vital organs in X-ray images [16, 17]. The CAD systems are mainly relying on the rapid development of computer technology such as graphical processing units (GPUs) to run the medical image processing algorithms, including image enhancement, organ and/or tumor segmentation, and interventional navigation tasks [18-20]. Now, artificial intelligence techniques such as machine learning and deep learning become the core of advanced CAD systems in many medical applications; for example, pulmonary diseases [21, 22], cardiology [23, 24], and brain surgery [25, 26].

Deep learning techniques showed in the last years promising results to accomplish radiological tasks by automatic analyzing multimodal medical images [27-29]. Deep convolutional neural networks (DCNNs) are one of the powerful deep learning architectures and have been widely applied in many practical applications such as pattern recognition and image classification in an intuitive way [30]. DCNNs are able to handle four manners as follow [31]: 1) training the neural network weights on very large available datasets; 2) fine-tuning the network weights of a pre-trained DCNN based on small datasets; 3) Applying unsupervised pre-training to initialize the network weights before putting DCNN models in an application; and 4) using pre-trained DCNN is also called an off-the-shelf CNN being used as a feature extractor. In previous studies, DCNNs have been exploited in X-ray image classification to successfully diagnose common chest diseases such as Tuberculosis screening [32] and mediastinal lymph nodes in CT images [33]. However, the application of deep learning techniques to identify and detect novel COVID-19 in X-ray is still very limited so far. Therefore, the aim of this study to propose a new framework of pre-trained deep learning classifiers; namely COVIDX-Net as an advanced tool to assist radiologists to automatically diagnose COVID-19 in X-ray images. In the following, the contributions of this paper are summarized.

- Building altogether deep learning models in a new framework (COVIDX-Net) to automatically assist the early diagnosis of patients with COVID-19 in an efficient manner.
- Achieving an empirical analysis of the proposed deep learning image classifiers in the task of classifying COVID-19 disease using conventional chest X-ray with lower cost than other imaging modalities like CT.
- Reporting comparative performances of different deep learning models with remarks to show the most accurate classification results of COVID-19 using a small X-ray image dataset.
- The proposed COVIDX-Net framework supports interdisciplinary researchers to continue developing advanced artificial intelligence techniques for CAD systems to fight the COVID-19 outbreak.

The rest of this paper is structured as follows. Section 2 gives a review on the state-of-the-art deep convolutional neural network models as image classifiers. Also, a detailed description of the COVIDX-Net framework is presented. Experimental results and comparative performance of the proposed deep learning classifiers are investigated and discussed in section 3. Finally, this study is concluded with the main prospects in section 4.

## 2. Methods

### 2.1 Deep learning image classifiers

In this section, we describe some of the existing state-of-the-art deep learning image classifiers that are required to accomplish the clinical purpose of the COVIDX-Net framework as follows.

A) **VGG19**: Visual Geometry Group Network (VGG) was developed based on the convolutional neural network architecture by Oxford Robotics Institute's Karen Simonyan and Andrew Zisserman [34]. It was addressed at the 2014 Large Scale Visual Recognition Challenge (ILSVRC2014). The VGGNet performed very well on the imageNet dataset. In order to have improved image extraction functionality, the VGGNet used smaller filters of 3×3, compared to AlexNet 11×11 filter. There are two versions of this deep network architecture; namely VGG16 and VGG19 have different depths and layers. VGG19 is deeper than VGG16. The number of parameters for VGG19, however, is larger and thus more expensive than VGG16 to train the network.

B) **DenseNet121:** The Dense Convolutional Network (DenseNet) have several compelling benefits: they lighten the vanishing-gradient problem, reinforce feature propagation, encourage feature reuse, and the number of parameters reduced substantially [35]. DenseNet121 is a Dense Net model which generated with 121 layers, the model was loaded with pre-trained weights from ImageNet database.

C) **InceptionV3:** Inception network or GoogLeNet was 22-layer network and it won 2014 Image net challenge with 93.3% top-5 accuracy [36]. Later versions are referred as Inception VN where N is the version number so inceptionV1, inceptionV2 and inceptionV3. The Inception V3 network has several symmetrical and asymmetrical building blocks, where each block has several branches of convolutions, average pooling, max-pooling, concatenated, dropouts, and fully-connected layers.

*D) ResNetV2:* He *et al*. [37] developed the Residual Neural Network (ResNet) models by utilizing skip connections to jump over some network layers to achieve strong convergence behaviors. The improved version of ResNet is called ResNet-V2. Although the ResNet is similar to the VGGNet, it is approximately eight times deeper [38].

*E) Inception-ResNet-V2:* A convolutional neural network is 164 layers deep, combining the Inception architecture with residual connections. Inception-ResNet-V2 is a variation of InceptionV3 [39]. Inception-ResNet-V2 is trained on more than a million images from the ImageNet database.

*F) Xception:* The architecture of Xception model is a linear stack of depth wise separable convolution layers with residual connections to easily define and modify the deep network architecture [40]. The Xception is an enhancement of the Inception architecture that replaces regular inception modules with distinguishable depth convolutions.

*G) MobileNetV2:* Sandler *et al*. [41] proposed the MobileNetV2 model as a convolutional neural network architecture for machines with limited computing power, like smartphones. The MobileNets achieve this key advantage by reducing the number of learning parameters, and introducing the inverted-residuals-with-linear-bottleneck-blocks to greatly reduce the memory consumption. Moreover, the pre-trained implementation of Mobile NetV2 is widely available in many popular deep learning frameworks.

## 2.2 Proposed COVIDX-Net Description

We proposed a new deep learning framework for automatically identifying the status of COVID-19 in 2D conventional X-ray images. Fig. 1 depicts the overall workflow of our proposed COVIDX-Net based on seven different architectures of DCNNs; namely VGG19, DenseNet201, InceptionV3, ResNetV2, InceptionResNetV2, Xception, and MobileNetV2, as described above.

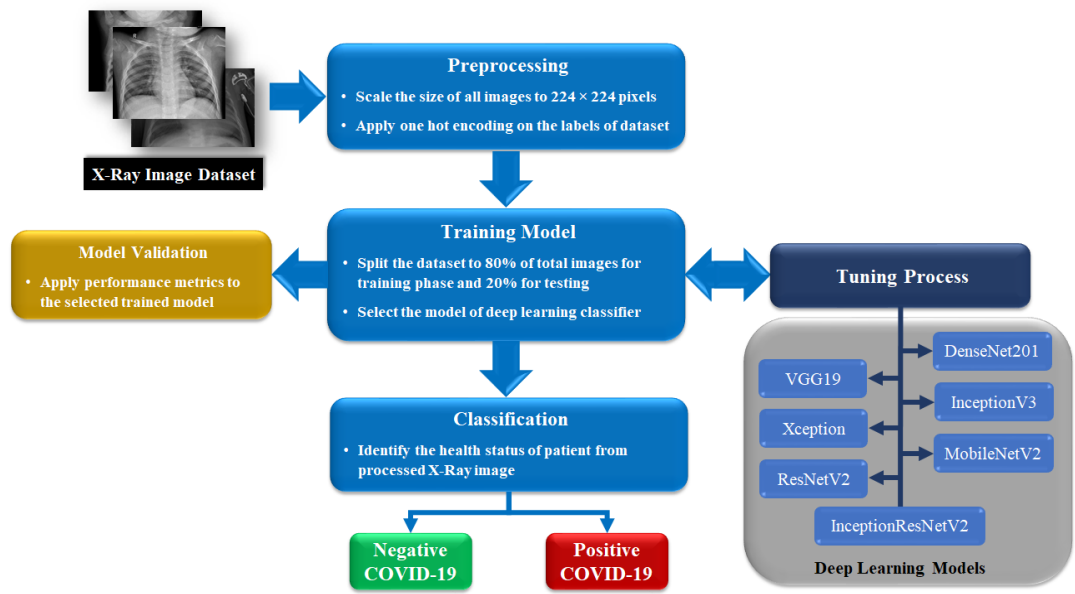

**Fig 1.** Workflow of proposed COVIDX-Net framework for classifying the COVID-19 status in X-Ray images.

The COVIDX-Net framework includes three main steps to accomplish the diagnostic procedure of novel Coronavirus, as follows.

- **Step#1: Preprocessing**

    All X-ray images have been collected in one dataset and loaded for scaling at a fixed size of 224 X 224 pixels to be suitable for further processing within the deep learning pipeline. One-hot encoding [42] is then applied on the labels of image data to indicate the case of positive COVID-19 or "not" for each image in the dataset.

- **Step#2: Training Model and Validation**

    In order to start the training phase of selected and/or tuned one of seven deep learning models, the preprocessed dataset is 80-20 split according to the Pareto principle. That means 20% of image data will be used for testing phase. Again, splitting 80% data will be used for constructing equal training and validation sets. Subsample random selections of training image data for the deep learning classifier, and then apply evaluation metrics to show the recorded performance on the validation set.

- **Step#3: Classification**

    In the final step of the proposed framework, the testing data is fed to the tuned deep learning classifier to categorize all the image patches into one of two cases: confirmed positive COVID-19 or normal case (negative COVID-19), as shown in Fig. 1. At the end of the workflow, the overall performance analysis for each deep learning classifier will be evaluated based on the metrics described in the following section.

## 2.3 Classification Performance Analysis

In order to evaluate the performance for each deep learning model in the COVIDX-Net, different metrics have been applied in this study to measure the true and/or misclassification of diagnosed COVID-19 in the tested X-ray images as follow. First, the cross validation estimator [43] was used and resulted in a confusion matrix as illustrated in Table 1. The confusion matrix has four expected outcomes as follows. True Positive (TP) is a number of anomalies and has been identified with the right diagnosis. True Negative (TN) is an incorrectly measured number of regular instances. False Positive (FP) is a collection of regular instances that are classified as an anomaly diagnosis FP. False Negative (FN) is a list of anomalies observed as an ordinary diagnosis.

**Table 1:** Confusion Matrix

|  | Predicted Positive | Predicted Negative |
|---|---|---|
| Actual Positive | True Positive (TP) | False Negative (FN) |
| Actual Negative | False Positive (FP) | True Negative (TN) |

After calculating the values of possible outcomes in the confusion matrix, the following performance metrics can be calculated.

A) *Accuracy:* Accuracy is the most important metric for the results of our deep learning classifiers, as given in (1). It is simply the summation of true positives and true negatives divided by the total values of confusion matrix components. The most reliable model is the best but it is important to ensure that there are symmetrical datasets with almost equal false positive values and false adverse values. Therefore, the above components of the confusion matrix must be calculated to assess the classification quality of our proposed COVIDX-Net framework.

$$\text{Accuracy}(\%) = \frac{TP+TN}{TP+FP+FN+TN} 100\% \quad (1)$$

B) *Precision:* Precision is represented in (2) to give relationship between the true positive predicted values and full positive predicted values.

$$\text{Precision} = \frac{TP}{TP+FP} \quad (2)$$

C) *Recall:* In (3), recall or sensitivity is the ratio between the true positive values of prediction and the summation of predicted true positive values and predicted false negative values.

$$\text{Recall} = \frac{TP}{TP+FN} \quad (3)$$

D) *F1-score:* F1-score is an overall measure of the model's accuracy that combines precision and recall, as represented in (4). F1-score is the twice of the ratio between the multiplication to the summation of precision and recall metrics.

$$\text{F1-score} = 2(\frac{\text{Precision} \times \text{Recall}}{\text{Precision} + \text{Recall}}) \quad (4)$$

## 3 Experiments
### 3.1 Dataset and Experimental setup

The public dataset of X-ray images used in this study for classifying negative and positive COVID-19 cases provided by Dr. Joseph Cohen[1] and Dr. Adrian Rosebrock[2]. The dataset includes 50 X-ray images, divided into two classes as 25 normal cases and 25 positive COVID-19 images. Fig. 2 shows a sample of normal and COVID-19 images extracted from the dataset. The X-ray images for confirmed COVID-19 disease show a pattern of ground-glass opacification with occasional consolidation in the patchy, peripheral, and bilateral areas [10]. The original size of tested images is ranging from 1112 x 624 to 2170 x 1953 pixels. For the experimental setup, all images were scaled to the size of 224 × 224 pixels. The COVIDX-Net framework including deep learning classifiers have been implemented using Python and the Keras package with TensorFlow2 [44] on Intel(R) Core(TM) i7-2.2 GHz processor. In addition, the experiments were executed using the graphical processing unit (GPU) NVIDIA GTX 1050 Ti and RAM with 4 GB and 16 GB, respectively.

---

[1] https://github.com/ieee8023/covid-chestxray-dataset
[2] https://www.pyimagesearch.com/category/medical/

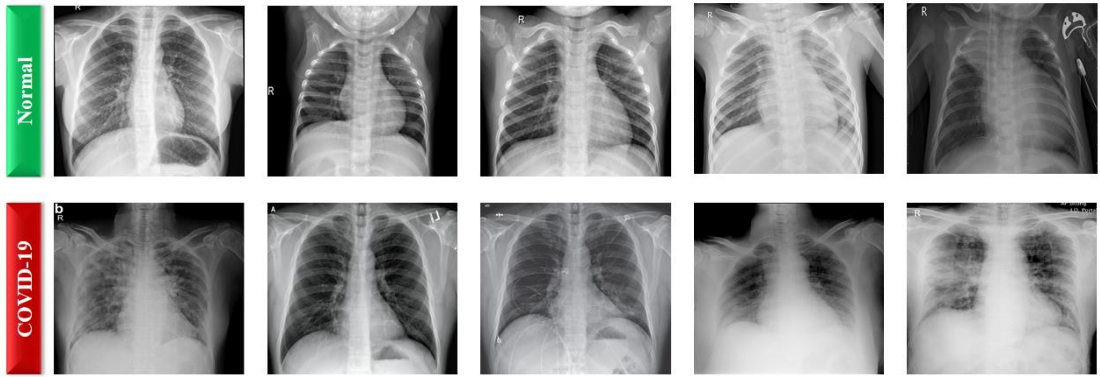

**Fig 2.** A sample of X-ray images dataset for normal cases (*first row*) and COVID-19 patients (*second row*)

### 3.2 Overall Performance Evaluation

For evaluating the performance of proposed deep learning classifiers, 80% of X-ray images including normal and diseased cases are randomly chosen for training, i.e. 40 images of the dataset. The training parameters for all DCNNs architectures in this study are: The learning rate = $e^{-3}$, the values of batch size and number of epochs are set to 7 and 50, respectively, to achieve the desired convergence with few iterations on this small X-ray image dataset, and also to avoid the degradation problem as possible. All deep network classifiers are trained using Stochastic Gradient Descent (SGD) because of its good converge and fast running time. Image data augmentation was not used in this study.

Table 2 illustrates the comparative computational times and the accuracy of tested deep learning image classifiers. The running times of all deep learning models are relatively short ranging from 390.0 to 2645.0 seconds because of using powerful capabilities of the GPU with a small X-ray image dataset. The resulted testing times of the COVIDX-Net models did not exceed 6 seconds on 10 tested images, as shown in Fig. 3. Among all tested classifiers, the accuracy of InceptionV3 model was the worst of 50 %, while the VGG19 and DenseNet201 models achieved the best values of accuracy (90%). Although the MobileNetV2 model showed a moderate value of accuracy (60%), it achieved the smallest computational times of 389.0 and 1.0 seconds for training and testing phases, respectively, as listed in Table. 2. In addition, the values of performance metrics of each deep learning classifier are presented in Table 3. The highest precision of deep learning classifier to detect only positive COVID-19 was achieved by ResNetV2, InceptionResNetV2, Xception, and MobileNetV2, but their corresponding performances were worst to classify the normal cases correctly. **Therefore, we recommend the VGG19 and DenseNet201 models to be applied for in the CAD systems to identify the health status of patients against the COVID-19 in X-ray images.**

Moreover, Fig. ٣ depicts the graphical performance evaluation of all trained deep learning classifiers with accuracy and cross-entropy loss (loss) in the training and validation step. The best scores of training and validation accuracy were achieved for VGG19 and DenseNet201 models, and the worst case is resulted by the InceptionV3, as illustrated also in Table 3. The resulted confusion matrices of all tested deep learning classifiers are depicted in Fig. ٤. Furthermore, our results added the Receiver Operating Characteristics (ROC) curves to verify the classification performances of each deep learning classifier by showing the true positive rate (TPR) against the false positive rate (FPR) to identify the positive COVID-19 cases in the tested X-ray images.

**Table 2:** Computational times and classification accuracy of all tested deep learning models of the COVIDX-Net on a GPU.

| Classifier | Training Time (seconds) | Testing Time (seconds) | Accuracy (%) |
|---|---|---|---|
| **VGG19** | 2641.00 | 4.00 | 90 |
| **DenseNet201** | 2122.00 | 6.00 | 90 |
| **ResNetV2** | 1086.00 | 2.00 | 70 |
| **InceptionV3** | 1121.00 | 2.00 | 50 |
| **InceptionResNetV2** | 1988.00 | 6.00 | 80 |
| **Xception** | 2035.00 | 3.00 | 80 |
| **MobileNetV2** | 389.00 | 1.00 | 60 |

**Table 3:** Comparative classification performance of deep learning models used in the COVIDX-Net

| Classifier | Patient Status | Precision | Recall | F1-score |
|---|---|---|---|---|
| **VGG19** | *COVID-19* | 0.83 | 1.00 | 0.91 |
| | *Normal* | 1.00 | 0.80 | 0.89 |
| **DenseNet201** | *COVID-19* | 0.83 | 1.00 | 0.91 |
| | *Normal* | 1.00 | 8.00 | 0.89 |
| **ResNetV2** | *COVID-19* | 1.00 | 0.40 | 0.57 |
| | *Normal* | 0.62 | 1.00 | 0.77 |
| **InceptionV3** | *COVID-19* | 0.00 | 0.00 | 0.00 |
| | *Normal* | 0.50 | 1.00 | 0.67 |
| **InceptionResNetV2** | *COVID-19* | 1.00 | 0.60 | 0.75 |
| | *Normal* | 0.71 | 1.00 | 0.83 |
| **Xception** | *COVID-19* | 1.00 | 0.60 | 0.75 |
| | *Normal* | 0.71 | 1.00 | 0.83 |
| **MobileNetV2** | *COVID-19* | 1.00 | 0.20 | 0.33 |
| | *Normal* | 0.56 | 1.00 | 0.71 |

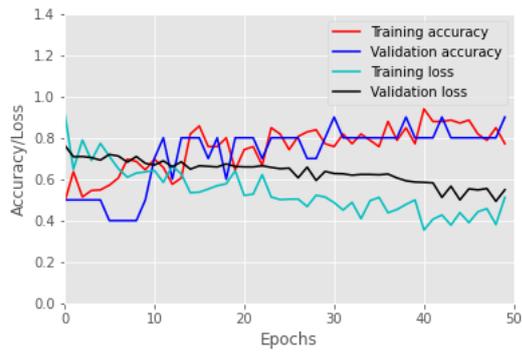
**(a)** VGG19

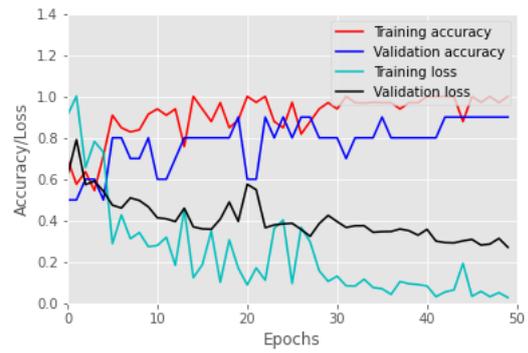
**(b)** DenseNet201

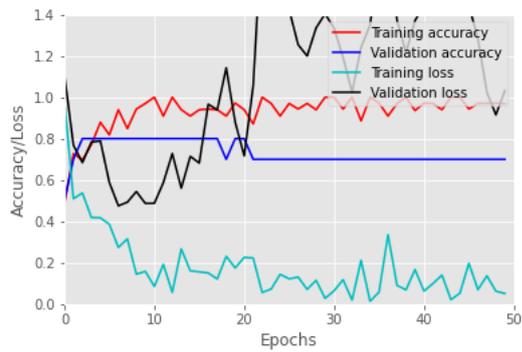
**(c)** ResNetV2

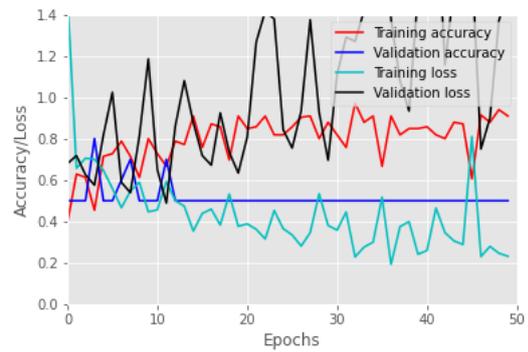
**(d)** InceptionV3

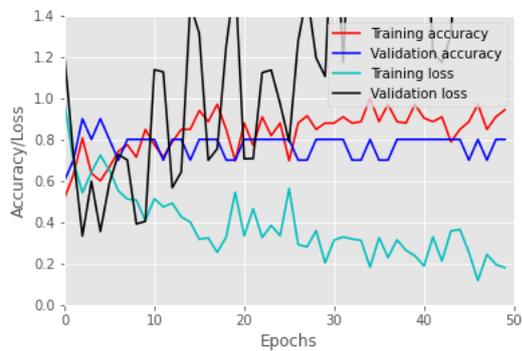
**(e)** InceptionResNetV2

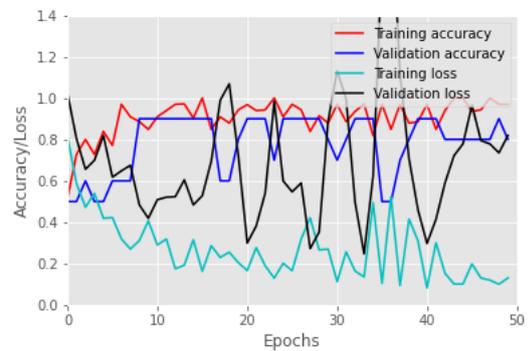
**(f)** Xception

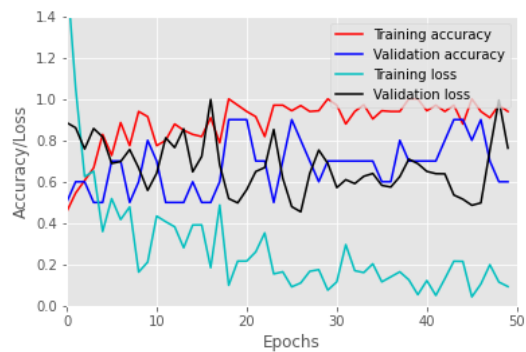
**(g)** MobileNetV2

**Fig 3.** Training loss and accuracy evaluation of all deep learning models in the COVIDX-Net

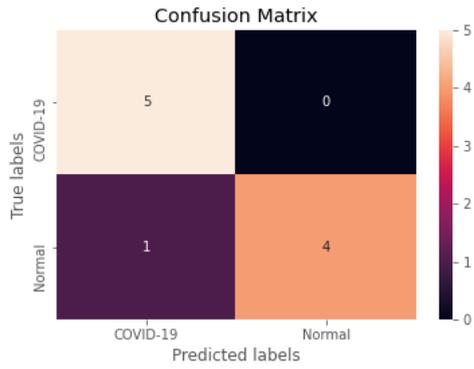

**(a)** VGG19

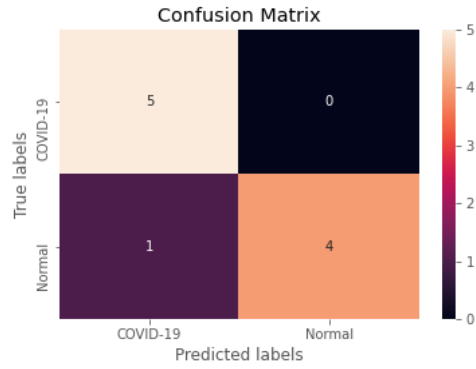

**(b)** DenseNet201

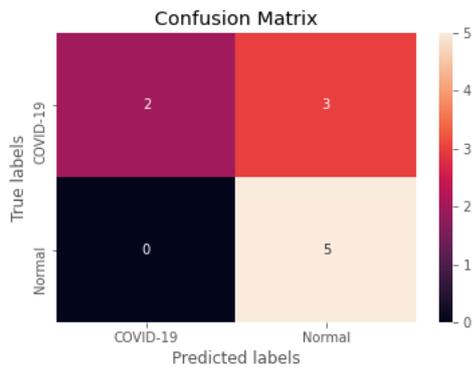

**(c)** ResNetV2

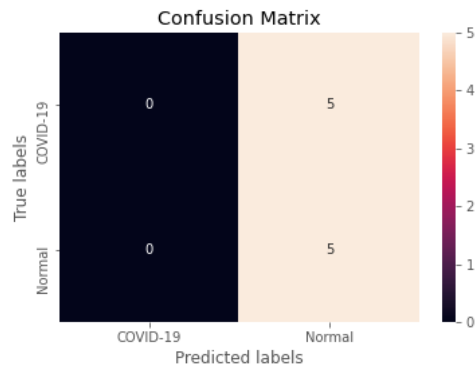

**(d)** InceptionV3

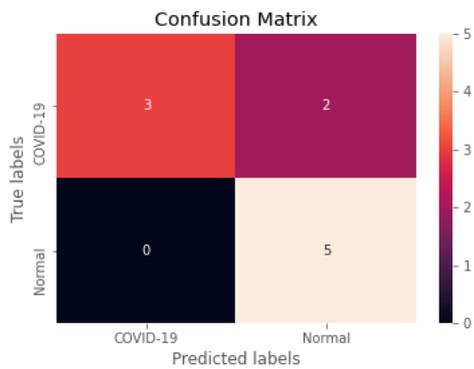

**(e)** InceptionResNetV2

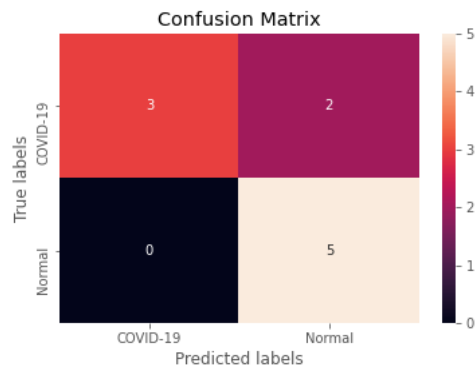

**(f)** Xception

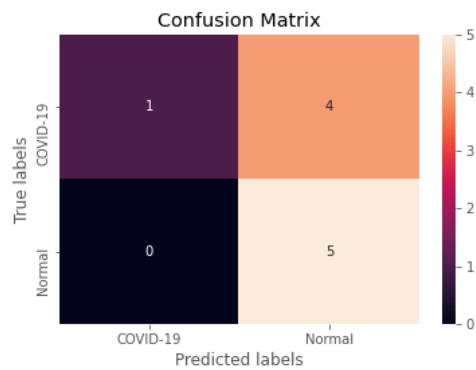

**(g)** MobileNetV2

**Fig 4**. Confusion matrix of all deep learning models in the COVIDX-Net

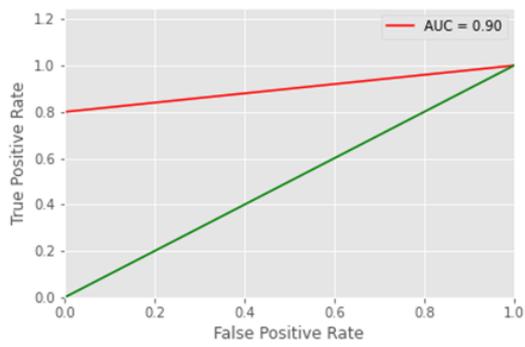
(a) VGG19

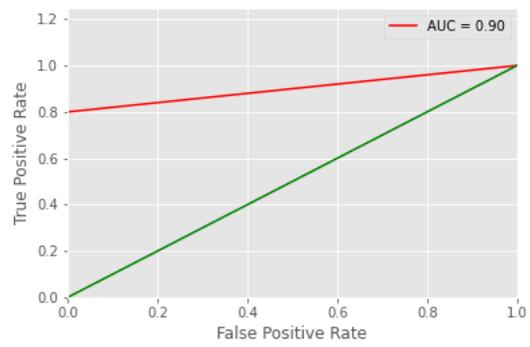
(b) DenseNet201

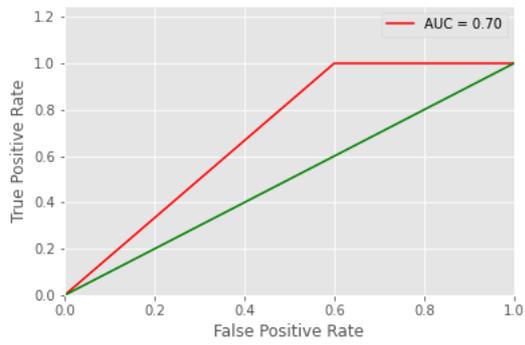
(c) ResNetV2

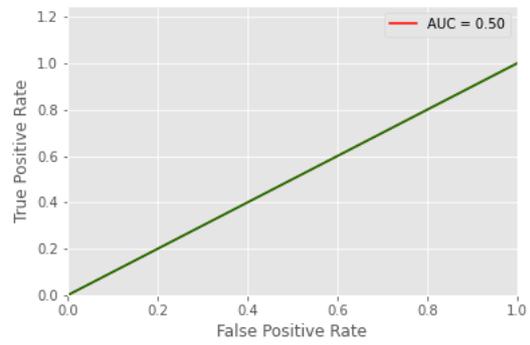
(d) InceptionV3

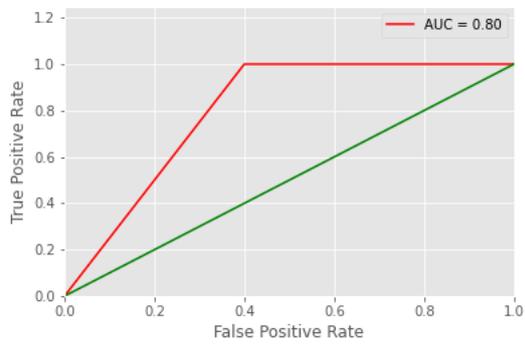
(e) InceptionResNetV2

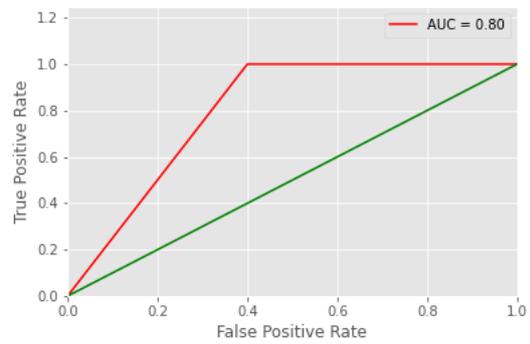
(f) Xception

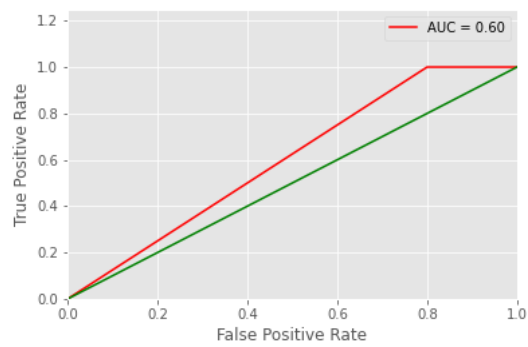
(g) MobileNetV2

**Fig 5** ROC curves of all deep learning models in the COVIDX-Net

## 4 Conclusions

Infectious COVID-19 disease shocked the world and is still threating the lives of billions of people. In this study, a new CVOIDX-Net framework has been proposed to automatically identify or confirm COVID-19 in 2-D X-ray images based on seven deep learning classifiers; namely VGG19, DenseNet121, ResNetV2, InceptionV3, InceptionResNetV2, Xception, and MobileNetV2. The results of our proposed COVIDX-Net verified that the best performance scores of deep learning classifiers are for the VGG19 and DenseNet201models. Furthermore, the performance of the MobileNetV2 can be further improved to be used on smart devices because of its high computational speed in the healthcare sector to assist the accurate cost-effective of COVID-19 in X-ray images and maybe later used for CT imaging modality in our future work of this highly important research topic.